# Is the Shroud of Turin in Relation to the Old Jerusalem Historical Earthquake?

Alberto Carpinteri, Giuseppe Lacidogna, Oscar Borla

*Politecnico di Torino, Department of Structural, Geotechnical and Building Engineering*
*Corso Duca degli Abruzzi 24 – 10129 Torino, Italy*

**Abstract.** Phillips and Hedges suggested, in the scientific magazine Nature (1989), that neutron radiation could be liable of a wrong radiocarbon dating, while proton radiation could be responsible of the Shroud body image formation. On the other hand, no plausible physical reason has been proposed so far to explain the radiation source origin, and its effects on the linen fibres. However, some recent studies, carried out by the first author and his Team at the Laboratory of Fracture Mechanics of the Politecnico di Torino, found that it is possible to generate neutron emissions from very brittle rock specimens in compression through piezonuclear fission reactions. Analogously, neutron flux increments, in correspondence to seismic activity, should be a result of the same reactions. A group of Russian scientists measured a neutron flux exceeding the background level by three orders of magnitude in correspondence to rather appreciable earthquakes (4th degree in Richter Scale). The authors consider the possibility that neutron emissions by earthquakes could have induced the image formation on Shroud linen fibres, trough thermal neutron capture by Nitrogen nuclei, and provided a wrong radiocarbon dating due to an increment in $C_6^{14}$ content. Let us consider that, although the calculated integral flux of $10^{13}$ neutrons per square centimetre is 10 times greater than the cancer therapy dose, nevertheless it is 100 times smaller than the lethal dose.

# 1 Introduction

After the first photographs of the Shroud, taken by Mr. Secondo Pia during the Exposition of 1898 in Turin [1], a widespread interest has been generated among scientists and curious to explain the image formation and to evaluate its dating. First results of radiocarbon analyses were published in 1988. They showed that the Shroud is at most 728 years old [2]. Later, some researchers have suggested that neutron radiation is liable of a wrong radiocarbon dating of the linen [3-5]. However, no plausible physical reason has been proposed so far to explain the radiation source origin.

Different documents in the literature attest the occurrence of disastrous earthquakes in the "Old Jerusalem" of 33 A.D., during the Christ's death [6-11]. On the other hand, recent neutron emission detections have led to consider the Earth's crust as a relevant source of neutron flux variations. Russian researchers measured neutron fluxes exceeding the background by three orders of magnitude in correspondence to seismic activity and rather appreciable earthquakes [12-17].

In this work, the authors consider that neutron emissions by earthquake –as for the conventional gadolinium-like neutron imaging technique– could have induced the image formation on Shroud linen fibres through thermal neutron capture by nitrogen nuclei, and provided a wrong radiocarbon dating due to an increment in $C_6^{14}$ content of carbon-14. Moreover, some recent studies, carried out by the authors at the Laboratory of Fracture Mechanics of the Politecnico di Torino, found that it is possible to generate neutron emissions from very brittle fracture of rock specimens in compression through piezonuclear phenomena. Neutron flux increase, in correspondence to seismic activity, should be a consequence of the same phenomena [18-20].

Starting from the first photographs of the Shroud, which highlighted a figure of a human body undraped with hands crossed (Fig. 1), a large debate on the cause that may have produced such an image has been conducted in the scientific community. The image seems to be formed with lights and shades reversed in a sort of negative photography. Vignon [1] asserts that the image was produced by radiographic action from the body which, according to ancient texts, was wrapped in a shroud impregnated with a mixture of oil and aloes. Other authors, instead, disapprove the observations of Vignon. In particular Waterhouse [21] affirms that, if a body were wrapped in a linen cloth, under the conditions stated in the Gospels, it would be impossible for such a detailed impression to be produced in the manner suggested by Vignon.

Further studies have focused on the Shroud dating, especially since 1986 [22], when the Roman Catholic Church declared that pieces of the Shroud of Turin had been sent to seven laboratories around the world, later reduced to only three [23], for radiocarbon dating. In 1988 Dickman [2] declares that, after weeks of rumour and speculation, the official carbon dating results for the Turin Shroud were released in Zurich. The results, also published in [24], provide evidence that the linen of the Shroud of Turin is medieval, dated between 1260 and 1390. Very recently, an exhaustive study on the statistical aspects of radiocarbon dating due to the heterogeneity caused by the division of the samples into subsamples, has been also published [25].

Phillips in the paper "Shroud irradiated with neutrons?" [3] supposes that the Shroud may have been irradiated with neutrons which would have changed some of the carbon nuclei to different isotopes by neutron capture. In particular, Phillips assumes that some $C_6^{14}$ nuclei could have generated from $C_6^{13}$, and that an integrated flux of $2\times10^{16}$ thermal neutrons $cm^{-2}$ could have produced an apparent carbon-dated age of just 670 years. However, in the reply to the same paper, Hedges [4] asserts that the integrated flux proposed by Phillips [3] is excessively high and that «including the neutron capture by nitrogen in the cloth, an integrated thermal neutron flux of $2 \times 10^{13}$ would be appropriate» for the apparent radiocarbon dating of the Shroud.

Also Rinaudo [26] evaluates that simultaneous fluxes of protons and neutrons could explain at the same time the imprint on the cloth (by protons) and the 13-century slip in time of the $C_6^{14}$ nuclei (by neutrons).

Fanti [27] confirms the hypothesis of Rinaudo stating that, at the same time of the neutron and proton emissions, also an electron emission could have generated the body image formation. Other Authors have recently introduced the hypothesis of radon emissions as a possible trigger of surface electrostatic discharges (ESD) and then of the image impression [28].

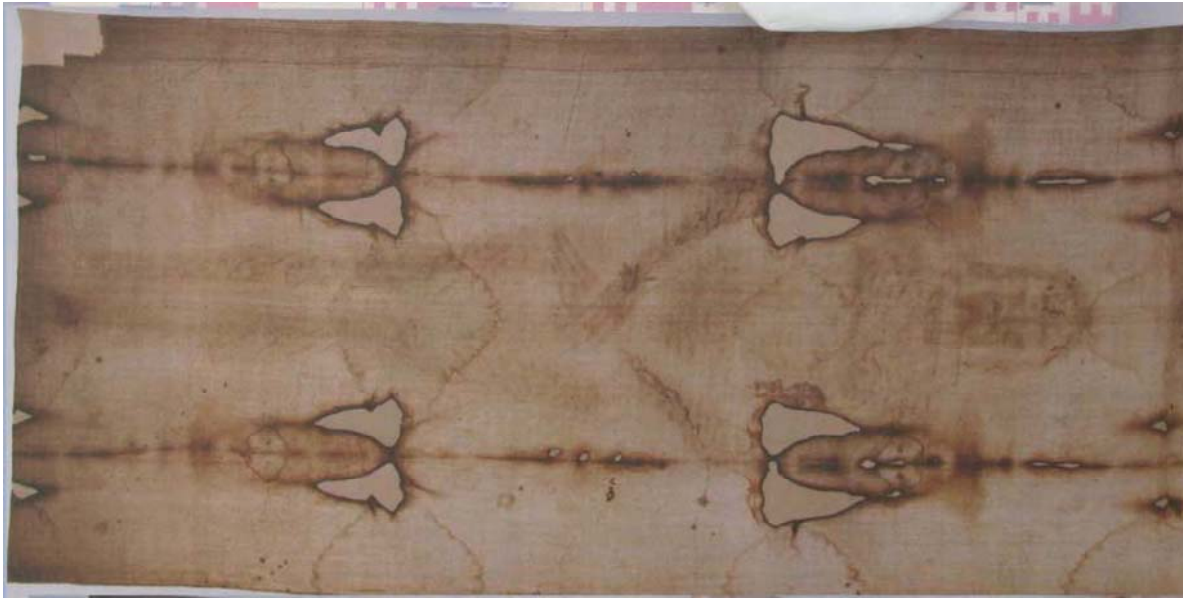

**Fig. 1:** Semitransparent front copy of the Turin Shroud (see Fanti, G., Basso, R. & Bianchini, G., Turin Shroud: Compatibility between a digitized body image and a computerized anthropomorphus manikin. Journal of Imaging and Technology 54(5), 050503 (1-8) (2010)).

**2 Earthquakes and Neutron Emissions from the Earth's Crust**

Scientific data of the historical earthquake occurred in 33 A.D. in the Jerusalem area are mentioned in the "Significant Earthquake Database" of the American Scientific Agency NOAA (National Oceanic and Atmospheric Administration) [29]. This database contains information on destructive earthquakes from 2150 B.C. to present days. The "Old Jerusalem" earthquake is classified as an average devastating seismic event, that it has also destroyed the City of Nisaea, the port of Megara, located at west of the Isthmus of Corinth [6]. It also would have involved to a total cost for the reconstruction that, if the current dollar amount of damages were listed, it would be between 1.0 and 5.0 million dollars.

In addition, if we assign the image imprinted on the Shroud to the Man who died during the Passover of 33 A.D., there are at least three documents in the literature attesting the occurrence of disastrous earthquakes during and after that event.

Within his chronicle in Greek language, a historian named Thallos, probably lived in Rome in the middle of the 1st century, has left mention of events occurred on the Christ's death day: the darkening of the sky and the happening of an earthquake [7,8]. The work of Thallos has been lost, but the quotation of the passage about Jesus had been inserted in the Chronographia of Sextus Julius Africanus, a Christian Palestinian author who died in Nicopolis around the 240 A.D.: «The most dreadful darkness fell over the whole world, the rocks were torn apart by an earthquake and much of Judaea and the rest of the land was torn down. Thallos calls this darkness an eclipse of the sun in the third book of his Histories, without reason it seems to me. For....how are we to believe that an eclipse happened when the moon was diametrically opposite the sun?».

Thallos, due to the quotation of Julius Africanus, is generally considered by historians as a witness to the early date of the gospel story of the "darkness" at the death of Christ: see Mark 15: 33; Luke 23: 44 and Mattew 27: 45 [9]. However, the interesting fact is that Julius Africanus criticizes Thallos, saying impossible that there was an eclipse on the day of Passover, which occurs in the full moon period, but he does not dispute that on the same day there was an earthquake.

On the other and, Matthew wrote that there was a strong earthquake at the moment of Christ's death: «When the centurion and those who were with him, keeping watch over Jesus, saw the earthquake and what took place, they were filled with awe and said, "Truly this was the Son of God!".» (Matthew 27: 54) [9]. He wrote that there was another even stronger earthquake at the time of the resurrection: «And behold, there was a great earthquake, for an angel of the Lord descended from heaven and came and rolled back the stone and sat on it. His appearance was like lightning, and his clothing white as snow. And for fear of him the guards trembled and became like dead men.» (Matthew 28:2-4) [9].

There is also the narrative of Joseph of Arimathea: «And, behold, after He had said this, Jesus gave up the ghost, on the day of the preparation, at the ninth hour. And there was darkness over all the earth; and from a great earthquake that happened, the sanctuary fell down, and the wing of the temple» (The Narrative of Joseph, Chapter 3, The good robber, 5) [10].

That event is also mentioned by Dante Alighieri, XXI Canto, Inferno, as the most violent earthquake that had ever shaken the Earth: « Poi disse a noi: "Più oltre andar per questo / iscoglio non si può, però che giace / tutto spezzato al fondo l'arco sesto. / E se l'andare avante pur vi piace, / andatevene su per questa grotta; / presso è un altro scoglio che via face. / Ier, più oltre cinqu'ore che quest'otta, / mille dugento con sessanta sei / anni compié che qui la via fu rotta"» (Inferno, XXI Canto:106-114) [11].

Since most scholars believe that the journey of Dante began on the anniversary of the Christ's death, during the Jubilee of 1300, the chronology goes back to 33 A.D., on the Friday when, according to tradition, Christ was put to death. Therefore, it was the earthquake after the Christ's death to cause disasters and crashes, including the Sanctuary of Jerusalem, and the wing of the Solomon's Temple [10].

Nevertheless, the results from historical studies have value for Earth scientists only when the information is converted into data representing epicentral location and magnitude of the events.
Modern scholars say that Jerusalem is situated relatively close to the active Dead Sea Fault zone. They accept the occurrence of the Resurrection earthquake, to which they assign the severity of a catastrophic event, characterized by a local magnitude ML = 8.2, as well as of another earthquake that took place in Bithynia, during the same period, that would have had even a greater magnitude [12].

Based on a detailed analysis of paleoearthquakes along the major active faults in the Earth's crust, some studies give evidence of their spatial and temporal distributions, as well as of their regional recurrent behaviour [30]. From these studies, it can be argued that a hypothetical earthquake of the 11th degree in the Richter scale magnitude may have a recurrence time of about 1000 years, as well as of about 100 years one of the 10th degree, and of about 10 years one of the 9th degree.

In the active faults of the Mediterranean basin and of the Middle East region, about one hundredth of the earthquakes recorded during long periods over the entire surface of the Earth take place, and their historical maximum intensity should be close to the 9th degree in the Richter scale [12]. In this case, an earthquake of the 9th degree may have a recurrence time of about 1000 years, as well as of about 100 years one of the 8th degree, and so on.

This last statistical remark would give further scientific value, as well as historical and archaeological importance, to the hypothesis that, in the "Old Jerusalem", there was a strong earthquake very close to the 9th degree in the Richter scale.

Recent neutron emission detections by Volodichev et al. [13], Kuzhevskij et al. [14,15], and Antonova et al. [16] have led to consider the Earth's crust as a relevant source of neutron flux variations. Neutron emissions measured in seismic areas of the Pamir region (4200 m asl) exceeded the usual neutron background "up to two orders of magnitude in correspondence to seismic activity and rather appreciable earthquakes, greater than or equal to the 4th degree in the Richter scale magnitude"[13].

On the other hand, it is important to note that the flux of atmospheric neutrons increases linearly starting from the top of the atmosphere up to a maximum value corresponding to an altitude of about 20 km. From this altitude it decreases exponentially up to the sea level, where is negligible (Pfotzer profile [31]). Considering the altitude dependence of neutron radiation, values about 10 times higher than the natural background at sea level are generally detected at 5000 m altitude. Therefore, the same earthquake occurring at sea level should produce a neutron flux up to 1000 times higher than the natural background. More recent neutron emission observations have been performed before the Sumatra earthquake of December 2004 [17]. Variations in thermal neutron measures were observed in different areas (Crimea, Kamchatka) a few days before that earthquake.

**3 Neutron Radiography and Neutron Imaging on Linen Fibres**

Considering the possibility that neutron flux could have induced appreciable effects on linen fibres of the Shroud, the authors have developed some hypotheses based on piezonuclear reactions [32]. In the following, it is briefly described the process of image formation induced by neutron radiation, occurred during the earthquake in the "Old Jerusalem" of 33 A.D., and the 13-century slip of time effects that could be produced on linen cloths.

Neutron radiography is an imaging technique that utilizes the transmission of neutron radiation to obtain a static picture of a given object [33]. The object under examination is placed in the path of the incident radiation. The transmitted and scattered neutrons "bring the visual information" of the object that is recorded by an appropriate imaging system.

Thanks to the high thermal neutron cross section of the gadolinium nucleus (~ 254000 barn), the most important detection reaction used in neutron imaging is:

$$\text{Gd}_{64}^{157} + \text{n}_{0}^{1} \rightarrow \text{Gd}_{64}^{158} + \text{gamma} + \text{conversion electrons}\,(8.5\,\text{MeV})$$

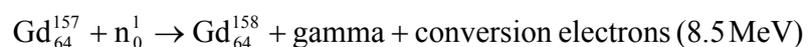

in which the converter material (gadolinium) captures neutrons and emits secondary charged particles that reproduce the irradiated object on neutron imaging plates (NIP). Usually, a thermal neutron flux of $10^5$ neutrons cm$^{-2}$ s$^{-1}$ is employed, with an irradiation time of few minutes for a total integrated flux of about $10^8$ neutrons cm$^{-2}$, with typical NIP enriched for more than 20% in weight of Gd$_2$O$_3$ [34].

Neutron imaging is a technique different, although complementary, to X-ray radiography. Whereas the X-rays are more sensitive to materials with rather high density, one of the advantages of neutron radiation is its ability to affect preferentially elements with low atomic numbers such as hydrogen, nitrogen, etc. [33]. For this reason, neutron rays give sharp images of biological soft tissue samples, whereas X-rays are more sensitive to tissues rich in calcium like bones.

The most important nuclear reaction of thermal neutrons on nitrogen nuclei is represented by:

$$\mathrm{N}^{14}_{7} + \mathrm{n}^{1}_{0} \rightarrow \mathrm{C}^{14}_{6} + \mathrm{H}^{1}_{1}$$

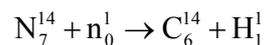

that is liable of radiocarbon formation also in the atmosphere.

For the chemical composition of linen fibres, a typical nitrogen concentration of 1000 p.p.m. could be supposed [4]. The nitrogen thermal neutron capture cross section is of about 1.83 barn. By comparison, $\mathrm{Gd}^{157}_{64}$ cross section is $10^5$ times greater than the nitrogen one. Neglecting the different concentrations in gadolinium and nitrogen, which mainly affect the image resolution, and taking into account only their cross sections, the thermal neutron flux necessary to nitrogen nuclei neutron imaging should be approximately equal to or higher than $10^{10}$ neutrons cm$^{-2}$ s$^{-1}$.

The hypothetical reaction induced by neutrons on nitrogen nuclei might have contributed on image formation also by means of proton radiation (as assumed by Rinaudo [26]), triggering chemical combustion reactions or oxidation processes on linen fibres. Usually, image formation from a neutron beam can be accomplished in a variety of ways by using a suitable conversion screen. Thus, through an etching process with a chemical reagent (like KOH or NaOH) and under appropriate lighting, the image will become visible. Similarly, in the case of linen, neutrons could have interacted with nitrogen nuclei, and the protons –produced as secondary particles– may have assumed the function of the reagent, triggering oxidation or combustion phenomena and making the image visible. Hypotheses and experimental confirmations that oxidative phenomena generated by earthquakes can provide 3D images on the linen clothes have recently been proposed by de Liso [35].

**4 Earthquake and Neutron Effects on the Shroud Radiocarbon Dating**

Taking into account the historical sources attesting the occurrence of a disastrous earthquake in 33 A.D., and assuming a hypothetical magnitude of the 9th degree in the Richter scale [12], it is possible to provide an evaluation of the consequent neutron flux.

The Richter scale is logarithmic (base 10). This means that, for each degree increasing on the Richter scale, the amplitude and the acceleration of the ground motion recorded by a seismograph increase by 10 times. From a displacement or acceleration viewpoint, the seismic event occurred in 33 A.D. may have been $10^5$ times more intense than the reference event of the 4th degree. On the other hand, from the energy viewpoint, it should have been $10^{10}$ times more intense than the same reference event [36].

Assuming a typical environmental thermal neutron flux background of about $10^{-3}$ $cm^{-2}\,s^{-1}$ at the sea level, in correspondence of earthquakes with a magnitude of the 4th degree, an average thermal neutron flux up to $10^0$ $cm^{-2}\,s^{-1}$ should be detected, that is 1000 times higher than the natural background [16], as previously calculated following the Pfotzer profile [31].

Thus, an earthquake of the 9th degree in the Richter scale could provide a thermal neutron flux ranging around $10^{10}$ $cm^{-2}\,s^{-1}$, if proportionality between released energy and neutron flux holds. A similar event could have produced chemical and/or nuclear reactions, contributing both to the image formation and to the $C_6^{14}$ increment in the linen fibres of the Shroud, if it had totally lasted for at least 15 minutes. In this way, an appropriate integrated thermal neutron flux of about $10^{13}$ neutrons $cm^{-2}$ is obtained, as exactly assumed by Hedges [4]. Let us consider that, although the calculated integral flux of $10^{13}$ neutrons per square centimetre is 10 times greater than the cancer therapy dose, nevertheless it is 100 times smaller than the lethal dose.

In confirmation of this assumption, one of the most powerful earthquakes, the so called "Greatest Chile Earthquake", occurred in Valdivia on May 22, 1960, had a complicated seismogram that lasted for at least 15 minutes [37]. Further information about the intensity and duration of this earthquake are reported in [38].

Neutron emissions have been detected not only at the Earth's crust scale, but also in laboratory compression experiments as shown in [18-20]. The tests have been carried out by using suitable $He^3$ and bubble type BD thermodynamic neutron detectors. The material employed for the tests was non-radioactive Luserna stone, a metamorphic rock deriving from a granitoid protolith. Neutron emissions from this material were found to be of about one order of magnitude higher than the ordinary natural background level at the time of the catastrophic failure. For basaltic rocks, the neutron flux achieved a level even two orders of magnitude higher than the background. In addition, a theoretical explanation is also provided in recent works by Widom et al. [39,40].

5 Conclusions

Recent neutron emission detections have led to consider the Earth's crust as a relevant source of neutron flux variations. Starting from these experimental evidences, the authors have considered the hypothesis that neutron emissions from a historical earthquake have led to appreciable effects on Shroud linen fibres. Considering the historical documents attesting the occurrence in the "Old Jerusalem" of a disastrous earthquake in 33 A.D., the authors assume that a seismic event with magnitude ranging from the 8th to the 9th degree in the Richter scale could have produced a thermal neutron flux of up to $10^{10}$ cm$^{-2}$ s$^{-1}$. Through thermal neutron capture by nitrogen nuclei, this event may have contributed both to the image formation, and to the increment in $C_6^{14}$ on linen fibres of the Shroud. Let us consider that, although the calculated integral flux of $10^{13}$ neutrons per square centimetre is 10 times greater than the cancer therapy dose, nevertheless it is 100 times smaller than the lethal dose.

Mechanical and chemical confirmations that high-frequency pressure waves, generated in the Earth's crust during earthquakes, can trigger neutron emissions were obtained at the Laboratory of Fracture Mechanics of the Politecnico di Torino, testing in compression very brittle rock specimens.